# A Deep Reinforcement Learning Framework for Rapid Diagnosis of Whole Slide Pathological Images

Tingting Zheng[1], Weixing chen[2], Shuqin Li[1], Hao Quan[1], Qun Bai[1], Tianhang Nan[1], Song Zheng[3], Xinghua Gao[3], Yue Zhao[1] and Xiaoyu Cui[1*]

[1] College of Medicine and Biological Information Engineering, Northeastern University, Shenyang, 110169, China
*Corresponding author:{cuixy}@bmie.neu.edu.cn
[2] Shenzhen College of Advanced Technology, University of the Chinese Academy of Sciences, Beijing 100049, China
[3] NHC Key Laboratory of Immunodermatology (China Medical University), No.155 Nanjing Bei Street, Heping District, Shenyang, Liaoning Province, 110001, China.

**Abstract.** Deep neural network is a research hotspot on histopathological image analysis, which can improve the efficiency and accuracy of diagnosis for pathologists or be used for disease screening. The whole slide pathological image can reach one gigapixel and contains abundant tissue feature information, which needs to be divided into a lot of patches in training and inference stages. This will lead to long convergence time and large memory consumption. Furthermore, well-annotated data sets are also in short supply in the field of digital pathology. Inspired by the pathologist's clinical diagnosis process, we propose a weakly supervised deep reinforcement learning framework, which can greatly reduce the time required for network inference. We use neural network to construct the search model and decision model of reinforcement learning agent respectively. The search model predicts the next action through the image features of different magnifications in the current field of view, and the decision model is used to return the predicted probability of the current field of view image. In addition, an expert-guided model is constructed by multi-instance learning, which not only provides rewards for search model, but also guides decision model learning by knowledge distillation method. Experimental results show that our proposed method can achieve fast inference and accurate prediction of whole slide image without any pixel-level annotations.

## 1 Introduction

In recent years, with the development of deep neural network (DNN), deep learning has become the mainstream method for analyzing pathological tissue images. Among malignant tumors with high morbidity and mortality, such as breast cancer [1], liver cancer [1], prostate cancer [3], lung cancer [4], gastric cancer [4], colon cancer [5], skin cancer [7]. DNN can excavate rich features of pathological tissue images, provide more objective histopathological image analysis, assist pathologists in making decisions and improve patient survival rate. The robustness and generalization performance of most DNN models depends on the availability of accurately labeled data [1,3,4]. However, whole slide pathology images (WSIs) have very high resolution, which makes pixel-level labeling difficult and expensive.

In order to solve this problem, researchers have proposed methods of weak supervision [1] or unsupervised learning [5], which only rely on the characteristics of data to provide supervised information without pixel-level annotation. However, training the model through WSIs requires hundreds of hours and huge storage space on the GPU [8]. In addition, due to the limitation of computer memory, the input image resolution of DNN is limited. When dealing with WSIs, image down sampling or dividing WSIs into small patches is generally adopted. But low-resolution images are not conducive to learning detailed features, especially for pathological tissue images, whose small changes may point to different disease outcomes [9]. Patch-based image processing of WSIs will also produce redundant data, which not only interferes with the learning of disease features, but also increases the time of model training.

During the paraffin section diagnosis, pathologists tend to use a low-power microscope to quickly scan the suspicious area, then switch to a high-power microscope to further identify tissue types, and finally give a comprehensive diagnosis [12]. However, the sequential nature of this visual diagnostic process is ignored by most existing DNN models, which always focus on distinguishing the difference of image features in the current state of a certain location. Recently, Xu et al. [10,10] formulated the pathologist's selective visual diagnosis as a Markov decision process (MDP) problem, and achieving 98% accuracy in 50% training time compared with the previous hard attention method. This is a meaningful attempt, but it does not fully reflect the decision making search behavior of pathologists, and the rewards are determined only by the predicted probability at a specific image resolution, so the sequential decision-making advantage of MDP is not fully utilized.

In this study, we propose a new weakly supervised Deep reinforcement learning framework (WSDRL) that decomposes the training process of DNN into two parts: disease area search behavior (Search Agent, SeAgent) and feature recognition behavior (Decision Agent, DeAgent). A multi-scale SeAgent is constructed to imitate the disease sequence search process of pathologists: "saccade under low magnification image, decision under high magnification image". A teacher-student model is built that uses pathologists' prior knowledge for self-guidance, and then identifies abnormal and normal tissue. The student model is constructed based on convolutional neural network, which imitating the task of disease feature recognition by pathologists during a diagnosis; while the teacher model is developed based on multi-instance learning (MIL), which simulating the prior knowledge that pathologists have mastered before diagnosis. The teacher model transfers the benign and malignant pathological tissue features in the prior knowledge to the student model through knowledge distillation (KL) [13], and then completes the classification task. The contributions of this work can be summarized in three aspects:

1. The WSDRL is proposed for computer aided diagnosis of WSIs, which can mimic the pathologist's clinical diagnostic process and has the advantage of high precision, low memory consumption and fast inference.
2. A multi-scale strategy search model is designed, which uses the teacher model based on multi-instance learning to feedback the rewards of the environment. The model can assist in setting the reward function in reinforcement learning and improve the search performance of agents.
3. A multi-instance learning teacher model (MILTM) is constructed to simulate the prior knowledge of pathologists, which can train the network to learn pathological image features and distinguish disease types through knowledge distillation.

## 2    Methods

The flow chart of our method is shown in Fig. 1. The details are described below.

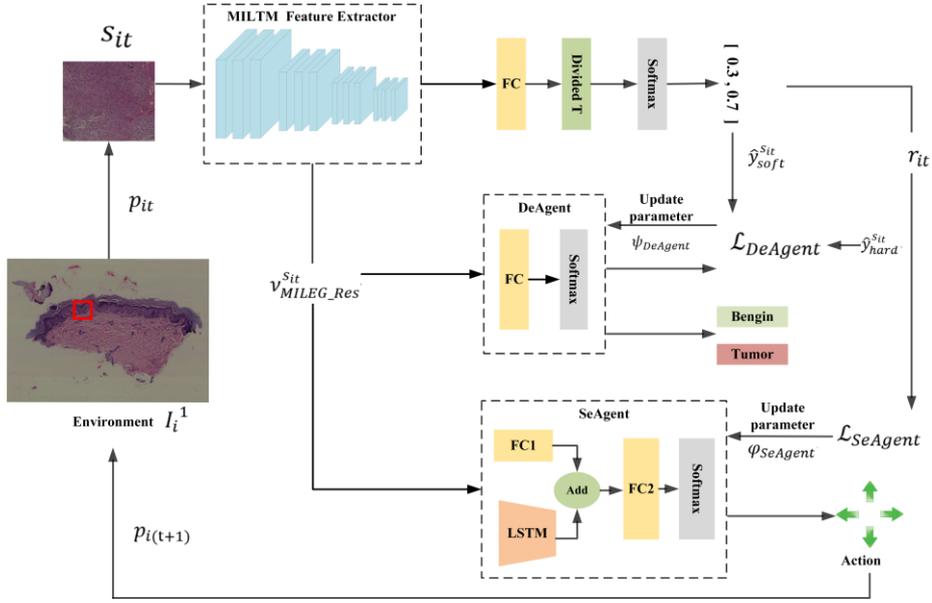

**Fig. 1.** Flow chart of weakly supervised deep reinforcement learning.

### 2.1    Problem Definition of Reinforcement Learning

In this study, the diagnostic process of pathologists is analogous to the interaction between agents and the external environment. The non-background tissue area extracted from WSI under $e = 1$ is taken as the external environment $I_i^1 \in \mathbb{R}^{W \times H \times C}$. $i \in N_i$ is interpreted for the $N_i$ WSIs. W, H, and C are width, length, and number of channels, respectively. $e \in \{1,2\}$, 1 and 2 represent WSI corresponding to 5x and 10x respectively. The basic ideas of MDP are used for modeling, $M =$ $\{State, Action, Reward, \gamma\}$, where $\gamma$ is the discount factor that controls the importance of future and current rewards.

**State Design.** State represents a finite state set $S_i = \{s_{it}\}$, and $s_{it} = \{I_i^1(p_{it})\}^{224 \times 224 \times 3}$ represents the state of the SeAgent at time t. $p_{it}$ is the position of SeAgent at time t. $t \epsilon T$, $T = \frac{(W \times H)}{step^2}$ represents the maximum number of times that is executed on each WSI for SeAgent, and step = 224 denotes the step size of SeAgent.

**Action Design.** Action is a limited set of actions $A_i = \{a_{it}\}$, $a_{it} \in \{0,1,2,3\}$. 0, 1, 2, and 3 represent the four actions of up, down, left, and right, respectively. According to the step and $a_{it}$, the $p_{it}$ and $s_{it}$ of SeAgent are obtained. when the $s_{it}$ obtained in the $I_i^1$ is judged as a tumor through MILTM, SeAgent will perform the zoom action and obtain the state $s'_{it}$ with $e = 2$ at the corresponding position.

**Reward Design.** Reward is the reward function that drives the action of SeAgent to select the maximum reward value. With Eq. (1), The reward is designed to encourage SeAgent to complete the task of disease diagnosis efficiently and quickly. If the $s_{it}$ is identified as a tumor by MILTM, it will feed back a "+1" reward to SeAgent, otherwise, a "-1" reward will be given. In addition, if $s'_{it}$ is also identified as a tumor by MILTM, the reward back to SeAgent is enlarged by a factor of 3. The rewards are generated immediately after each action.

$$r_{it} = \begin{cases} -1 & if\ \hat{y}_{MILEG}^1 = 0 \\ +1 & if\ \hat{y}_{MILEG}^1 = 1\ and\ \hat{y}_{MILEG}^2 = 0 \\ +3 & if\ \hat{y}_{MILEG}^1 = 1\ and\ \hat{y}_{MILEG}^2 = 1 \end{cases} \quad (1)$$

$\hat{y}_{MILEG}^1$ and $\hat{y}_{MILEG}^2$ represent the results predicted by $\mathcal{F}_{MILEG}^1$ and $\mathcal{F}_{MILEG}^2$ at states $s_{it}$ and $s'_{it}$, respectively.

$$\hat{y}_{MILEG}^1 = argmax\ (\mathcal{F}_{MILEG}^1(s_{it}\ |\ \theta_{MILEG}^1)) \quad (2)$$

$$\hat{y}_{MILEG}^2 = argmax(\mathcal{F}_{MILEG}^2(s'_{it}\ |\ \theta_{MILEG}^2)) \quad (3)$$

Calculating the maximum cumulative return by Eq. (4) to guide SeAgent to optimize. $\gamma^{t-1}$ is the weighting factor of reward $r_{it}$ at time t.

$$R_i = \sum_{t=1}^{T} \gamma^{t-1} r_{it} \quad (4)$$

### 2.2 Weak Supervised Deep Reinforcement Learning

**Multi-instance Teacher Model.** The MILTM is constructed through 34-layer residual convolution neural network (ResNet34). There are $N_i$ WSIs in dataset $\{(X_i, Y_i)\ |\ i = 1,2,\ldots,N_i\}$, and $Y_i$ are the labels of the corresponding $X_i$, $Y_i \in \{Benign, Tumor\}$. We design a bag dataset $B^e = \{B_i^e\ |\ i = 1,2,\ldots,N_i\}$. A WSI $X_i$ is regarded as a bag $B_i^e =$ $\{x_{ij}^e | j = 1,2,...,M_j\}$. $x_{ij}^e$ represents that when the magnification of $X_i$ is $e$. $M_j$ images with a size of $224 \times 224$ are randomly selected; For each $B_i^e$, the model select the top $K = 8$ instances with the highest response probability $x_{ij}^e$ and assign the corresponding label $Y_i$, then construct a new dataset $\hat{B}_i^e = \{(x_{ik}^e, y_{ik}) | k = 1,2,3,...,8\}$, $y_{ik} = Y_i$, with the selected instances and corresponding labels. MILTM is trained on dataset $\hat{B}_i^e$ using cross-entropy loss (CE). When $e$ takes 5x or 10x, these data are constructed as two datasets $\hat{B}_i^1$, $\hat{B}_i^2$, then used to train two MILTMs. Their parameters and network functions are denoted as $\mathcal{F}_{MILEG}^1$, $\theta_{MILEG}^1$ and $\mathcal{F}_{MILEG}^2$, $\theta_{MILEG}^2$, respectively.

**Search Agent.** The SeAgent is designed using Long Short-Term Memory (LSTM) combined with fully connected layers (FC). By optimizing the parameter $\varphi_{SeAgent}$, SeAgent can quickly find tumor areas. The input of SeAgent is the feature vector $v_{MILEG\_Res}^{s_{it}}$ of the $s_{it}$. $v_{MILEG\_Res}^{s_{it}}$ represents the feature vector of $s_{it}$ output by MILTM. Using pre-trained MILTM to extract state feature vector not only shortens the time used by SeAgent for feature extraction, but also improves the accuracy of SeAgent. Compared with convolutional neural networks and standard recurrent neural networks, LSTM is very suitable for the classification, processing and prediction of long-term time series data. In addition, it can solve the gradient disappearance phenomenon caused by the increase of network depth. $v_{MILEG\_Res}^{s_{it}}$ is sent into LSTM and FC respectively, and Eq. (5) is used to fuse feature $v_{SeAgent\_LSTM}^{s_{it}}$ output by LSTM and feature $v_{SeAgent\_FC1}^{s_{it}}$ output by FC. The $v_{fusion}^{s_{it}}$ go through FC $\mathcal{F}_{SeAgent}^{FC2}$ and the softamx function, finally output the next action of SeAgent $\pi_{\phi_{SeAgent}}(s_{it}, a_{it})$.

$$v_{fusion}^{s_{it}} = v_{SeAgent\_LSTM}^{s_{it}} + v_{SeAgent\_FC1}^{s_{it}} \tag{5}$$

$$\begin{aligned}\pi_{\phi_{SeAgent}}(a_{it} | s_{it}) &= P_{\phi_{SeAgent}}(a_{it} | s_{it}, \varphi_{SeAgent}) \\ &= softmax(\mathcal{F}_{SeAgent}^{FC2}(v_{fusion}^{s_{it}} | \varphi_{SeAgen\_FC2}))\end{aligned} \tag{6}$$

Since SeAgate is not differentiable, the policy gradient is used to optimize the parameter $\varphi_{SeAgent}$ to learn the optimal search policy $\pi_{\phi_{SeAgent}}(a_{it} | s_{it})$, and then maximize the expectation of $R_i$. Its defined objective function is:

$$\mathcal{L}_{SeAgent} = \mathbb{E}[\sum_{t=1}^{T} \gamma^{t-1} r_{it}] = \sum_{t=1}^{T} P_{\phi_{SeAgent}}(a_{it} | s_{it}, \varphi_{SeAgent}) \gamma^{t-1} r_{it} \tag{7}$$

Every time SeAgent completes a task, it updates the parameters of the search network. The equation for calculating the gradient of the objective function $\mathcal{L}_{SeAgent}$ is:

$$\nabla_{\phi_{SeAgent}} \mathcal{L}_{SeAgent} = \sum_{t=1}^{T} \gamma^{t-1} r_{it} \nabla_{\phi_{SeAgent}} \log(\pi_{\phi_{SeAgent}}(a_{it} | s_{it})) \tag{8}$$

**Decision network.** The DeAgent is built with Single FC. Optimize the parameter $\psi_{DeAgent}$ of the FC $\mathcal{F}_{DeAgent}$ to achieve accurate classification of $s_{it}$ feature vector $v_{MILEG\_Res}^{s_{it}}$. Since WSIs contain a large number of non-tumor tissue regions, it is difficult to optimize DeAgent using only the global labels of WSIs and the CE without pixel-level annotations. In addition, the label of the patch obtained by the SeAgent at each step is not necessarily consistent with the global label. Inspired by the teacher-student model, the CE is combined with KL [14] to guide DeAgent learning. After $s_{it}$ is input into the MILTM model, the patch information $z_{MILEG}^{s_{it}}$ is smoothed as a soft label $\hat{y}_{soft}^{s_{it}(H)}$, $\hat{y}_{soft}^{s_{it}(H)} = softmax\left(\frac{z_{MILEG}^{s_{it}}}{H}\right)$, where $H$ is the temperature coefficient, and it is used to adjust the information contained in the soft label $\hat{y}_{soft}^{s_{it}(H)}$. The label $Y_i$ of the tissue slice $X_i$ corresponding to $s_{it}$ is one-hot encoded as the hard label $\hat{y}_{hard}^{s_{it}}$. Kullback-Leibler divergence (KLdiv) [15] is used as a soft loss to measure the difference between $\hat{y}_{soft}^{s_{it}}$ and DeAgent output information $\hat{y}_{DeAgent}^{s_{it}(H)}$, $\hat{y}_{DeAgent}^{s_{it}(H)} = softmax\left(\frac{z_{DeAgent}^{s_{it}}}{H}\right)$. $z_{DeAgent}^{s_{it}}$ is the FC output of DeAgent. CE is used as a hard loss to calculate the error between DeAgent's prediction $\hat{y}_{DeAgent}^{s_{it}}$ and the ground truth $\hat{y}_{hard}^{s_{it}}$. The hyperparameter $\varepsilon$ is set to balance the proportion of KL and CE in the total loss. A larger $\varepsilon$ weighting coefficient indicates that the induced DeAgent relies on the contribution of MILTM, which is crucial for guiding DeAgent to recognize benign and malignant tissues in the early stages of training. Its specific equation is:

$$\mathcal{L}_{DeAgent} = \varepsilon * H^2 * KLdiv\left(\hat{y}_{soft}^{s_{it}(H)}, \hat{y}_{DeAgent}^{s_{it}(H)}\right) + (1-\varepsilon) * CE\left(\hat{y}_{hard}^{s_{it}}, \hat{y}_{DeAgent}^{s_{it}}\right) \quad (9)$$

## 3 Experiments

### 3.1 Dataset

The experiment uses a private data set containing 212 samples, including 105 patients with melanoma and 107 patients with benign nevus. All WSIs were reviewed by three dermatologists to confirm the diagnosis. The data set is randomly divided into 6: 2: 2 as training, verification and test set. Digital dermatopathological tissue slices were scanned under 40x (0.23 μm/pixel) using the HAMAMATSU-NanoZoomer 2.0-RS scanner to obtain data in ndpi format. Before the experiment began, the images were standardized. The image size for the input network is set to 224×224.

### 3.2 Experimental Setup and Evaluation Metric

The model is implemented through the PyTorch library and trained and inferred on a server configured with two TITAN_XP GPU. When MILTM is trained, the model is optimized by using cross entropy loss function and Adam. Learning rate = 0.0001, batch size = 128, epoch = 150. When training WSDRL, the same environment $I_i^1$, loop 150. SeAgent is optimized by gradient strategy, $\gamma = 0.9$, and DeAgent is optimized by combining KLdiv and CE. $H = 5$, $\varepsilon = 0.8$. Recall, precision, specificity, F1 Score (F1), Accuracy (ACC) are used to evaluate the model.

### 3.3 Results

**Results Comparison.** A series of MIL based models are used to compare and evaluate the effectiveness of our methods. As shown in Table 1, our method can improve the diagnostic accuracy on the basis of MIL method, especially for sensitivity, which indicates that our search strategy can find valuable diagnostic regions and can fuse image features of different magnification.

**Table 1.** Result comparison. VGG19/VGG16/ResNet50/ResNet34 represents a multi-instance model based on different networks. 10x/5x represents the magnification of WSI.

|     | Methods | Precision | Specificity | F1 | ACC |
| --- | --- | --- | --- | --- | --- |
| 10x | MIL-VGG19 | 0.833±0.178 | 0.800±0.118 | 0.897±0.063 | 0.887±0.095 |
|     | MIL-VGG16 | 0.827±0.060 | 0.787±0.085 | 0.904±0.035 | 0.893±0.043 |
|     | MIL-ResNet50 | 0.853±0.047 | 0.825±0.065 | **0.920±0.027** | 0.912±0.032 |
|     | MIL-ResNet34 | **0.863±0.071** | **0.837±0.085** | 0.919±0.026 | **0.912±0.032** |
| 5x  | MIL-VGG19 | 0.873±0.138 | 0.833±0.202 | 0.928±0.082 | 0.916±0.101 |
|     | MIL-VGG16 | 0.858±0.087 | 0.825±0.126 | 0.921±0.052 | 0.907±0.057 |
|     | MIL-ResNet50 | 0.815±0.077 | 0.775±0.106 | 0.900±0.043 | 0.887±0.053 |
|     | MIL-ResNet34 | **0.885±0.061** | **0.867±0.076** | **0.938±0.034** | **0.930±0.038** |
|     | **Ours** | **0.950±0.034** | **0.953±0.031** | **0.927±0.009** | **0.932±0.005** |

**Inference Runtime Comparison.** In order to evaluate the time taken by different methods in the inference stage, experiments were conducted on the test set. Set the batch size to 1. When the accuracy of all methods is around 90% (Table 1), the CPU running time of this method is only about 20% of that of other methods (Table 2). It shows that the method can greatly shorten the inference time while obtaining high accuracy.

**Table 2.** Performance Comparison. VGG19/VGG16/ResNet50/ResNet34 represents a multi-instance model based on different networks. 10x/5x represents the magnification of WSI.

|     | Methods | Run Time (s) |
| --- | --- | --- |
| 10x | MIL-VGG19 | 4568.14±23.388 |
|     | MIL-VGG16 | 4126.89±12.637 |
|     | MIL-ResNet50 | 3264.01±11.521 |
|     | MIL-ResNet34 | 3282.13±2.916 |
| 5x  | MIL-VGG19 | 4250.66±3.449 |
|     | MIL-VGG16 | 3843.99±2.754 |
|     | MIL-ResNet50 | 2814.29±2.355 |
|     | MIL-ResNet34 | 2831.50±2.613 |
|     | **Ours** | **723.68±11.970** |

### 3.4 Ablation study

In order to evaluate the effectiveness of each module in our method, the ablation experiments are performed (Table 1). When comparing with DeAgent without KLdiv or CE as loss functions, the Recall and ACC are improved by 10% and 8%, 6% and 4% respectively, which show that it is difficult to optimize DeAgent using only the global labels of WSIs and the CE without pixel-level annotations. When removing SeAgent zoom operation (only at 5x, the agents search for suspicious regions and diagnoses), the Recall, F1 and ACC decreased by about 3%, 5% and 6%, respectively, indicating that the combination of 5x and 10x to design reward can better encourage SeAgent to find suspicious areas. Finally, when compared to removing LSTM and FC, F1 improved by 4% and 2%, respectively, which show that the combination of LSTM and FC can make better use of the past feature information and improve the accuracy of the network.

**Table 3.** Ablation study. o-De-kldiv/CE removes KLdiv or CE from DeAgent. o-Se-FC1/LSTM removes the first FC or LSTM layer of SeAgent. o-se-FC1&10x /LSTM&10x removes the first FC and zoom action or LSTM and zoom action of SeAgent.

| Methods | Recall | Precision | Specificity | F1 | ACC |
|---------|--------|-----------|-------------|-----|-----|
| o-De-KLdiv | 0.807±0.099 | 0.912±0.010 | 0.923±0.017 | 0.854±0.060 | 0.874±0.046 |
| o-De-CE | 0.820±0.044 | 0.941±0.050 | 0.945±0.038 | 0.876±0.042 | 0.892±0.035 |
| o-Se-LSTM | 0.897±0.044 | 0.875±0.041 | 0.888±0.038 | 0.886±0.038 | 0.892±0.036 |
| o-Se-FC1 | 0.871±0.044 | 0.946±0.046 | 0.950±0.039 | 0.906±0.023 | 0.916±0.020 |
| o-Se-10x | 0.875±0.096 | 0.900±0.135 | 0.875±0.189 | 0.878±0.052 | 0.875±0.065 |
| o-Se-FC1&10x | 0.846±0.063 | 0.937±0.042 | 0.948±0.033 | 0.888±0.041 | 0.901±0.034 |
| o-Se-LSTM&10x | 0.794±0.088 | 0.942±0.048 | 0.947±0.038 | 0.871±0.049 | 0.892±0.036 |
| **Ours** | **0.908±0.042** | **0.950±0.034** | **0.953±0.031** | **0.927±0.009** | **0.932±0.005** |

## 4  Conclusion

This study simulates the visual diagnosis process of pathologists on WSIs, and proposes a brand new weakly supervised deep reinforcement learning, which uses MIL to construct an expert model to guide RL agents to learn. Different from other deep reinforcement learning methods, this study regarded the pathological diagnosis process as a complete MDP. Only rely on global labels, quickly find suspicious areas, reduce unnecessary patches processing, achieve high precision and greatly shorten the inference time of the model. In addition, this method can be combined with other models to provide more methods of deep reinforcement learning with short inference time and strong interpretability in the field of computer pathology. In the future, we will implement our proposed approach on more models and data sets, and continue to explore the application of deep reinforcement learning in computer-aided diagnosis.